\documentclass[%
 reprint,
 longbibliography,
 amsmath,amssymb,
 aps,
floatfix,
]{revtex4-2}

\usepackage{graphicx}
\usepackage{dcolumn}
\usepackage[dvipsnames]{xcolor}
\usepackage{bm}

\usepackage{empheq,etoolbox}
\begin{document}


\title{Stable vortex in Bose-Einstein condensate dark matter}


\author{Y.O. Nikolaieva$^{1}$, A.O. Olashyn$^{1,2}$, Y.I. Kuriatnikov$^{1,3}$, S.I. Vilchynskii$^{1,4}$, and A.I. Yakimenko$^{1}$}
 \affiliation{ $^1$ Department of Physics, Taras Shevchenko National University of Kyiv, 64/13, Volodymyrska Street, Kyiv 01601, Ukraine}
  \affiliation{ $^2$ Instituut-Lorentz, Universiteit Leiden, P.O. Box 9506, 2300 RA Leiden, The Netherlands}
\affiliation{ $^3$ Atominstitut, TU Wien, Stadionallee 2, 1020 Vienna,
Austria}
\affiliation{$^4$ Institute of Physics, Laboratory for Particle 
6
 Physics and Cosmology (LPPC), \'{E}cole Polytechnique F\'{e}d\'{e}rale de Lausanne (EPFL), CH-1015 Lausanne, Switzerland}





\begin{abstract}
The nature of dark matter (DM) is one of the most fascinating unresolved challenges of modern physics. One of the perspective hypotheses suggests that DM consists of ultralight bosonic particles in the state of Bose-Einstein condensate (BEC). The superfluid nature of BEC must dramatically affect the properties of DM matter including quantization of the angular momentum. Angular momentum quantum in the form of a vortex line is expected to produce a considerable impact on the luminous matter in galaxies including density distribution and rotation curves. We investigate the evolution of spinning DM cloud with typical galactic halo mass and radius. Analytically and numerically stationary vortex soliton states with different topological charges have been analyzed. It has been shown that while all multi-charged vortex states are unstable, a single-charged vortex soliton is extremely robust and survives during the lifetime of the Universe. 
\end{abstract}



\maketitle

\section{\label{sec:level1}Introduction}
Most of the recent cosmological models for galaxy structure are amount to show the galaxy as a luminous galactic baryon disk surrounded by a spherical galactic halo of so-called Dark Matter (DM). Different estimations \textcolor{black}{(such as direct detection, gamma ray detection,  microlensing measuring) \cite{2020,1996PhR...267..195J,2005PhR...405..279B,Schmaltz_2005,2007PhR...453...29H,2010ARA&A..48..495F}} 
give DM roughly 95$\%$ of the total mass of the galaxy.
The nature of DM remains one of the most exciting open questions in modern physics. 
Weakly Interacting Massive Particles (WIMPs) with a mass of O(100) GeV have been one of the leading DM candidates for a long time \cite{Arcadi_2018}. 
However, recent negative results for indirect detection \cite{Aprile_2018} 
and collider experiments \cite{Roszkowski_2018} 
cause strong motivation for developing alternative DM models.
Plenty of theoretical models have shown that considering sub-GeV DM has some advantages. One of such theories suggests considering DM as ultralight bosonic particles in the state of Bose-Einstein condensate (BEC) \cite{Hui_2017}. 
These bosonic particles can interact via gravity and probably via weak interaction. Both of these interactions are extremely insignificant (for instance, a mass of axions (one of the possible candidates for DM)  is estimated to be in the range $10^{-25}-10^{-2}$ eV \cite{Klaer_2017}; 
$s$-wave scattering length corresponds to the two-particle interaction varies in the very wide range $10^{-100}-10^{-1}$ fm). However, at the galactic and astrophysical scales, the self-gravitating BEC may form stable structures in the form of the galactic halo and astrophysical cold dark matter (CDM)  structures (Bose-stars) \cite{Kolb_1993}. 
Despite the extremely small mass of these bosonic particles, their gravity force dominates the Universe. 

The natural question arises: what is the physical mechanism that stands behind the bosonic DM self-stabilisation?  There are two most probable mechanisms to compensate the gravitational self-attraction and prevent collapse: (i) The quantum pressure that occurs whenever the condensate density is inhomogeneous. (ii) Interparticle repulsive interaction, which in mean-field approach leads to nonlinear self-induced potential proportional to condensate density. Each of these two mechanisms can stabilize the self-gravitating BEC and may lead to a formation of soliton-like stationary in time spatial structures. These nonlinear self-organized structures are well known in various physical systems. 

The hypotheses that dark matter structures of astrophysical and galactic scales can be treated as self-gravitating BEC composed of extremely light bosonic particles have been developed for decades.
%
Bose stars as lumps of Bose-Einstein condensates bound by self-gravity were proposed over 50 years ago \cite{PhysRev.187.1767,Tkachev:1986tr}. The stability of such objects has been studied previously numerically in a non-relativistic regime \cite{2015PhRvD..91d4041M}. Also, the formation of Bose stars \cite{2018PhRvL.121o1301L} and their collapse \cite{2017PhRvL.118a1301L} have been already studied.  The collisional dynamics of stable solitary waves in the Schrödinger–Poisson equation have been discussed in Ref.\cite{2016PDU....12...50P}.
Also have been analytically studied basic properties of self-gravitating BEC in a harmonic trap with Hartree–Fock method and compared with numerical calculations \cite{2001CQGra..18.1513J}.


The superfluid nature of BEC can dramatically affect the properties of DM including the formation of quantum vortices and quantization of angular momentum. There have been studied possible eﬀects of subgalactic vortices in the DM on the rotation velocity curves of virialized galaxies with standard DM halo proﬁles \cite{734543}. In Ref.\cite{2010PhRvD..82f4042K} exact solution for a single axisymmetric vortex has been found analytically in the Thomas-Fermi regime.
Conditions of vortex formation in
galactic halos composed of BEC DM have been discussed analytically in  Ref.\cite{2012MNRAS.422..135R}. 
The case of rigid rotation and its impact on BEC DM with and without self-interaction has been examined in Ref.\cite{2014PhRvD..89f3507G}. Rigid slow rotation of BEC DM has been investigated analytically in Tomas-Fermi limit in Ref.\cite{2018EPJC...78..346Z}.
Also, stability and dynamical properties of slowly rotating gravitationally self-bound BEC have been studied in \cite{2014PhRvD..90j3526K}.
It is remarkable that gravity-like attractive nonlocal interaction has been extensively studied in the context of atomic BEC (see recent review article \cite{2020PhyD..40332301P}). 
Similar to nonlocal optical media \cite{PRE05,PRE06} and BECs with long-range dipole-dipole \cite{2009PhyS...79c5305L} interactions stable spinning solitons and azumthons have been predicted in atomic BECs with gravity-like attractive interactions  \cite{PhysRevA.81.063617}.

In the present work, we consider CDM  of galactic scales. Our main goal is a consistent analysis of vortex structures in self-gravitating BECs. 
We address the following questions: (i) is it possible to balance such condensate in a state with nonzero angular momentum? (ii) is such a spinning superfluid CDM halo stable, and (iii) how the vortex structure manifests itself in observable properties of the luminous matter of the galaxy? 




Our paper is organized as follows. In Sec.\ref{sec:model} we define the model which we use to investigate the system. In Sec.\ref{sec:stationary_sol} we investigate the stationary solutions in two approaches: using variational analysis and using numerical modelling. In Sec.\ref{sec:dynamics} the dynamics of the vortex structures in self-gravitating BEC with typical galactic mass is investigated numerically.  We summarize our results in the concluding Sec.\ref{sec:conclusions}.

\section{\label{sec:model}Model}
At the zero-temperature limit, all the bosons condense into the same quantum ground state and the system is described by
single condensate wave function  $\Psi(\mathbf{r},t)$.
In the mean-field  approximation, the dynamics  of self-gravitating BEC  of $N$ weakly interacting bosons  with mass $m$  is described by the Gross-Pitaevskii-Poisson (GPP) system of equations \cite{Quantum}:
\begin{eqnarray}
\label{GPP1}
i \hbar\frac{\partial \Psi}{\partial t} = \left(-\frac{\hbar^2}{2m}\nabla^2 +g N|\Psi|^2+m\Phi \right)\Psi \nonumber \\
\nabla^2\Phi=4\pi GmN|\Psi|^2,
\end{eqnarray}
where $\Psi(\mathbf{r},t)$ is a complex wave function of the condensate with normalization condition $\int |\Psi|^2 d\textbf{r}=1$, $g=4\pi\hbar^2a_s/m$ is the coupling strength that corresponds to the repulsive two-particle interaction, 
$a_s$ is the $s$-wave scattering length, 
$\mathbf{r}=(x,y,z)$ - spacial coordinates, $t$ is time, $\Phi(\mathbf{r},t)$ is the gravitational potential, $G$ is the gravitational constant. The density field can be written as:
\begin{equation}
\label{densfield}
 \rho =mN|\Psi|^2,
\end{equation} 
where $mN=M$ is the total mass of the galaxy's halo. The total energy associated with the GPP system  can be written as 
\begin{equation}
\label{Etot}
E=\Theta +U+W,
\end{equation}
the kinetic energy
\begin{equation}\label{TKE}
   \Theta=\frac{N\hbar^2}{2m}\int |\nabla\Psi|^2d\mathbf{r}, 
\end{equation}
the internal energy
\begin{equation}
    \label{IE}
U = \frac{2\pi a_s \hbar^2}{m^3}\int \rho^2d\mathbf{r},    
\end{equation}
and the gravitational potential energy of interaction
\begin{equation}
    \label{GEI}
    W =\frac{1}{2}\int \rho\Phi d\mathbf{r}.
\end{equation}

In terms of dimensionless units $[\mathbf{r}\rightarrow\mathbf{r}/L_*, t\rightarrow\Omega_* t, E\rightarrow E/\epsilon]$ the system Eq.(\ref{GPP1}) can
be written in dimensionless form,
\begin{eqnarray}
\label{eq:GPEP_num}
\nonumber
 i\frac{\partial\Psi(\mathbf{r},t)}{\partial t} = \left(-\frac{1}{2}\nabla^2 +\Phi(\mathbf{r},t) +|\Psi(\mathbf{r},t)|^2 \right) \Psi(\mathbf{r},t),\\
 \nabla^2 \Phi(\mathbf{r},t)=|\Psi(\mathbf{r},t)|^2, 
\end{eqnarray}
where $L_*=\lambda_C (m_\textrm{Pl}/m)\sqrt{\lambda/8\pi}$,  $\Omega_*=c\lambda_C/L_*^2$,  $ \epsilon = (\hbar^2/4\pi m_{Pl}\lambda_{C}^2)\left(8\pi/\lambda\right)^{3/2}$, $m_{\textrm{Pl}}=\sqrt{\hbar c/G}$ is the Planck mass, $\lambda/8\pi=a_s/\lambda_{C}$ is the self-interaction constant, and $\lambda_{C}=\hbar/mc$ is the Compton wavelength of the bosons. With the new dimensionless gravitational potential $\Phi\rightarrow(L_*/\lambda_C)^2\Phi/c^2$ and the wave function $\Psi\rightarrow (\lambda/8\pi)\left(m_\textrm{Pl}/m\right)^2\sqrt{4\pi G M}(\hbar/mc^2)\Psi$. From now on in our paper, we use dimensionless variables.

  Finally, the normalization condition in dimensionless units:
\begin{equation}\label{eq:normalized_dimless}
  \int|\Psi|^2d\textbf{r}=4\pi\frac{M}{m_\textrm{Pl}} \sqrt{\frac{\lambda}{8\pi}}=N_0.   
\end{equation}

Note that the system Eqs.(\ref{eq:GPEP_num}) is invariant under following transformation: $t = \lambda_*^2\hat{t}$, $x = \lambda_* \hat{ x}$, $y = \lambda_* \hat{y}$, $z = \lambda_* \hat{z}$, $\psi = \lambda_*^{-2}\hat{\psi}$, $\Phi = \lambda_*^{-2}\hat{\Phi}$, $g = \lambda_*^2\hat{g}$, where $\lambda_*>0$. This scaling invariance allowed us to scale-out the coupling constant $g=1$ in Eqs.(\ref{eq:GPEP_num}).

The initial dimensional GPP system includes three crucial physical parameters: particle mass $m$, self-interaction constant $\lambda/8\pi$ (or, equivalently coupling constant $g$), and total mass of the system $M$ (or, equivalently, the total number of particles $N$). With these parameters, the system is fully described in our model.  In order to be specific, we fix two of them, which leaves the third one variable. 

We determine the particle mass as $m = 3\cdot10^{-24}\text{eV}$ and the self-interaction constant as $\lambda/8\pi=5.62\cdot10^{-98}$. This allows us to vary total halo mass $M$ and, as a result, describe different DM halos. When we fix the self-interaction constant $\lambda/8\pi$, the normalization constant $N_0$ is determined by the total halo mass. Our choice of determination for particle mass $m$ and self-interaction constant $\lambda/8\pi$ is described 
below.

It is important to note that all parameters of the dimensionless system Eqs. (\ref{eq:GPEP_num}) are completely
described by the normalization constant $N_0$. Thus, our results can be straightforwardly generalized
for arbitrary particle mass, self-interaction constant, and total halo mass. 

\section{\label{sec:stationary_sol}Stationary solutions}
Here we consider stationary localized DM structures, which
may appear as the result of a balance between gravitational contraction and two repulsive interactions: quantum pressure and nonlinear interaction. We study steady states of the GPP system for the condensate with topological charge $s$ 
Such a system has spherical symmetry for the fundamental soliton $s=0$ and cylindrical symmetry for vortex solitons $s>0$. 
Fundamental solitons  ($s=0$) have been already studied in variational approach \cite{Quantum}. In our work, we generalize these results for spinning $s$-charged vortex states. 

 The stationary solution wave function can be written in the following form: 
\begin{equation*}
\Psi(\mathbf{r},t) = \psi(r)e^{{-i\mu t}},    
\end{equation*}
where $\mu$ is a chemical potential and $\psi(r)$ is a radial profile of the wave function.
In case of cylindrical symmetry, the spatial part can be written as:
\begin{equation*}
  \psi(r_{\perp},\theta,z) = \chi(r_{\perp},z)e^{{is\theta}},
\end{equation*}
where $r_{\perp} = \sqrt{x^2+y^2}$.

\begin{figure}[htpb]
	\includegraphics[width=3.4in]{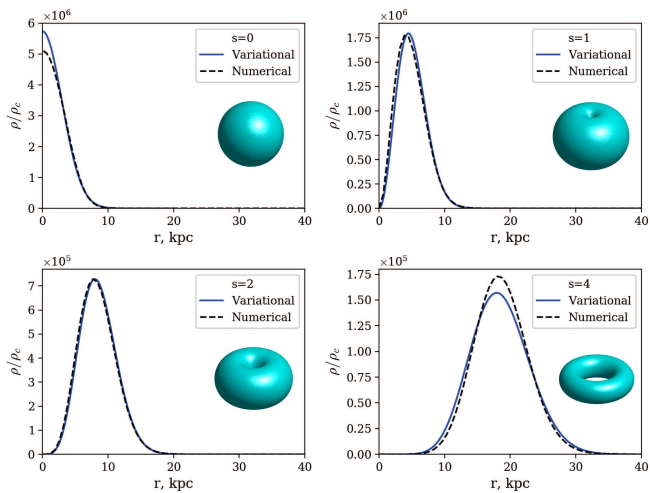}
	\caption{Condensate density profile $\rho/\rho_c$ in $(x,y,z=0)$ plain  as a function of radial coordinate in kpc for the halo of mass $M = 3\cdot10^{11}M_{\odot}$ with different topological charges: $s=0,1,2,4$.  Here  $\rho_{c}=8.5 \cdot 10^{-27}\textrm{kg}/{\textrm{m}^3}$ is the critical cosmological density. Solid blue line and dashed black line correspond to the variational analysis and numerical modeling respectively. The insets represent corresponding 3D density isosurfaces (surfaces of constant density) in cyan color.
	}
	\label{FIG:Rho}
\end{figure}

In our work, we use two methods: variational analysis and numerical modelling. The results obtained for topological charges $s=0,1,2,3,4$ with both methods show good agreement with each other (see Fig. \ref{FIG:Rho}). 
\begin{figure}[htpb]
	\includegraphics[width=3.4in]{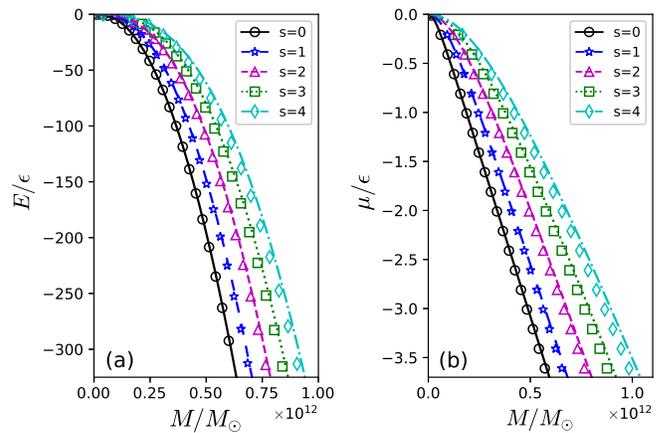}
	\caption{(a) The total energy $E$  and (b) chemical potential $\mu$ (in units of  $\epsilon =7.05\cdot10^{50}\text{J}$) of the stationary solitonic and vortex stuctures as functions of the halo mass $M$  (in units of $M_{\odot}\times 10^{12}$). Lines correspond to the variational analysis results while points correspond to the numerical modeling results for different topological charges: black solid line and  circles for $s=0$; blue long-dashed line and stars for $s=1$; magenta dashed line triangles for $s=2$; green dotted line and squares for $s=3$; cyan dash-dot line  and diamonds for $s=4$. }
	\label{FIG:Energy_mu}
\end{figure}

\begin{figure}[htpb]
	\includegraphics[width=3.4in]{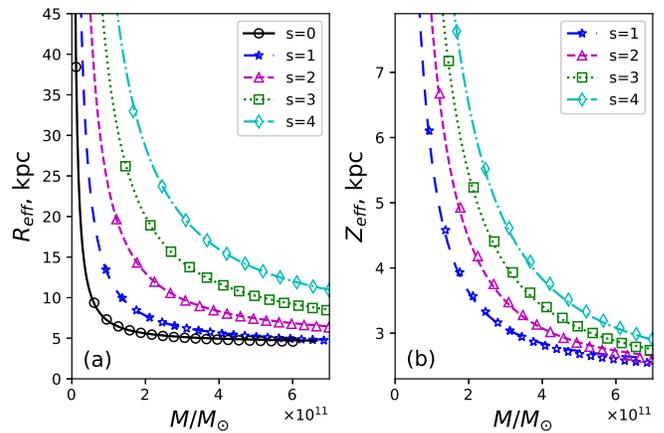}
	\caption{Effective radius (a) and effective height (b) (def. Eqs.(\ref{eq:R0_eff}), (\ref{eq:R_eff}), (\ref{eq:Z_eff})) as a functions of halo mass $M$ (in units of $M_{\odot}\times 10^{11}$). Lines correspond to the variational analysis results while points correspond to the numerical modeling results for different topological charges: black solid line and  circles for $s=0$; blue long-dashed line and stars for $s=1$; magenta dashed line triangles for $s=2$; green dotted line and squares for $s=3$; cyan dash-dot line  and diamonds for $s=4$.}
	\label{FIG:RZ_eff}
\end{figure}
\subsection{Variational analysis}
In order to gain a deeper insight into the properties of stationary solutions of GPP, we introduce a simple analytical variational analysis with a trial function of the form
\begin{equation}
\label{ansatz}
\psi(r_{\perp},\theta,z)= A \left(\frac{r_\perp}{R}\right)^{s} e^{-\frac{r_\perp^2}{2R^2}
-\frac{z^2}{2(R\eta)^2}+i s\theta},
\end{equation}
where $R$ and $\eta$ are variational parameters, $r_{\perp} = \sqrt{x^2+y^2}$. The constant A is defined by normalization condition Eq.(\ref{eq:normalized_dimless}):
\begin{equation*}
    A = \sqrt{ \frac{N_0}{\pi^{3/2}\eta R^3s!}}.    
\end{equation*}
Let us calculate the total energy functional Eq.(\ref{Etot}) using ansatz Eq.(\ref{ansatz}). The kinetic energy Eq.(\ref{TKE}) in dimensionless units
\begin{equation*}\label{eq:E_kinetic}
    \Theta=\frac{1}{2}\int |\nabla\psi|^2d\mathbf{r} = \frac{N_0(1+2\eta^2(1+s)) }{4R^2\eta^2},
\end{equation*}
the  internal energy Eq.(\ref{IE}) in dimensionless units
\begin{equation*}
    U = \frac{1}{2}\int |\psi|^4d\mathbf{r} = \frac{N_0^2\Gamma(s+1/2)}{4\sqrt{2}\pi^2R^3\eta\Gamma(s+1)},
\end{equation*}
 the gravitational potential energy of interaction  Eq.(\ref{GEI}) in dimensionless units
 \begin{equation*}
     W =\frac{1}{2}\int |\psi|^2\Phi d\mathbf{r},
 \end{equation*}
 the last integral can be calculated in Fourier space:
 \begin{equation*}
     W = \frac{1}{2}\int\mathcal{F}[|\psi|^2]\mathcal{F}[\Phi]d\mathbf{k}.
 \end{equation*}
 
 Making Fourier transform of the first equation in Eq.(\ref{eq:GPEP_num}), one can obtain the gravitational potential in Fourier space:
 \begin{equation}\label{eq:Fourier_Phi}
     \mathcal{F}[\Phi] = - \frac{1}{k^2}\mathcal{F}[|\psi|^2].
 \end{equation}
 Therefore,
  \begin{equation}
  \label{eq:Fourier_W}
     W = - \frac{1}{2} \int\frac{d\mathbf{k}}{k^2}{(\mathcal{F}[|\psi|^{2}}])^2,
 \end{equation}
and the Fourier transform of $|\psi|^2$:
\begin{equation}\label{eq:Fourier_rho}
    \mathcal{F}[|\psi|^2] = \frac{N_0}{(2\pi)^{3/2}}L_{s}(k_{\perp}^2R^2/{4})e^{{-\frac{k_z^2R^2\eta^2}{4}} - \frac{k_{\perp}^2R^2}{4}},
\end{equation}
 where $k_{\perp} =\sqrt{k_x^2+k_y^2}$ and $ L_{s}(x)$ is $s$-th order Laguerre polynomial.  
 
 Inserting this into Eq.(\ref{eq:Fourier_W}) and integrating over polar angle and $k_z$, one can obtain the following result for the gravitational potential energy:
 
 
 \begin{eqnarray*}
     W= - \frac{N_0^2}{8\pi}\int_0^\infty dk_{\perp}&& \text{Erfc}\left(\frac{k_{\perp}R\eta}{\sqrt{2}}\right)\\
     &&\times
     L_{s}^2\left(\frac{k_{\perp}^2R^2}{4}\right)e^{- \frac{k_{\perp}^2R^2(1-\eta^2)}{2}}.
 \end{eqnarray*}

Finally, the dimensionless total energy
\begin{equation}
\begin{split}
&E/\epsilon = \frac{N_0(1+2\eta^2(1+s)) }{4R^2\eta^2}+\frac{N_0^2\Gamma(s+1/2)}{4\sqrt{2}\pi^2R^3\eta\Gamma(s+1)}\\
   &- \frac{N_0^2}{8\pi R}\int_0^\infty \text{Erfc}(k_{*}\eta/\sqrt{2}) L_{s}^2(k_{*}^2/{4})e^{- \frac{k_{*}^2(1-\eta^2)}{2}}dk_{*}
\end{split}
\label{eq:dimless_energy}
\end{equation}

The next step of variational analysis is to find the minimum of the total energy in space of variational parameters $(R,\eta)$. This is done for the different topological charges s.

Let us discuss the choice of the BEC parameters and analysis of the fundamental soliton $s=0$.
This case corresponds to the spherical symmetry, therefore, there is only one variational parameter $R$ ($\eta=1$). Dimensionless total energy :
\begin{equation*}
    E/\epsilon =\frac{3}{4} \frac{N_0}{R^2} +  \frac{N_0^2}{4\sqrt{2}\pi^{3/2}R^3}  - \frac{N_0^2}{4\sqrt{2}\pi^{3/2}R}.
\end{equation*}

Energy minimum $dE/d R=0$ is attained at the point:
\begin{equation}\label{eq:mass_radius}
R = \frac{3\sqrt{2\pi^3}}{N_0}\left(1+\sqrt{1+\frac{N_0^2}{6\pi^3}}\right).
\end{equation}
Inserting this into Gaussian ansatz Eq.(\ref{ansatz}), one can find the density function, which is shown in the top left plot of the Fig. \ref{FIG:Rho} . Also, the energy and the chemical potential for this case are shown in Fig. \ref{FIG:Energy_mu}. In order to estimate the system spatial scales, it is useful to calculate the mean-squared radius which is called an effective radius. In this case 
\begin{equation}
    \label{eq:R0_eff}
    R_{\text{eff}}^2 = \frac{1}{N_0}\int |\psi|^2r^2d\textbf{r},
\end{equation}
where $r = \sqrt{x^2+y^2+z^2}$. The results for effective radius are shown in left plot in Fig. \ref{FIG:RZ_eff}.

The Eq.(\ref{eq:mass_radius}) is called "mass-radius relation" (because $N_0$ is proportional to mass) and  has been already achieved for GPP system in the variational analysis approach in \cite{Quantum}. Our interest here is, by using this relation, determine the physical parameters of the system - particle mass $m$ and self-interaction constant $\lambda/8\pi$.

 In order to determine them, we consider the following physical parameters for galactic halo: total mass $M = 3\cdot 10^{11}M_{\odot}$ and radius $R_\textrm{halo}=10$ kps $= 3.09\cdot 10^{20}$ m. These parameters are introduced in \cite{Quantum} as typical for DM condensate halo. 
 
 The next step is to calculate the radius, inside which the total mass of the halo is $0.99M$. This radius is called $R_{99}$ and in dimensionless units is calculated from the following equation:
 \begin{equation*}
     \int_0^{R_{99}}|\psi|^2 r^2d r=0.99\int_0^{\infty}|\psi|^2r^2dr.
 \end{equation*}
  Solving this equation, one can find:
 $
     R_{99} = 2.38R.
$
 Then, we fix the $R_{99}$ in physical units to be equal to the typical halo radius $R_\textrm{halo}$: 
$
    R_{99}\cdot L_* = R_\textrm{halo}
$
Using the mass-radius relation Eq.(\ref{eq:mass_radius}), the definition of the normalization constant $N_0$ Eq.(\ref{eq:normalized_dimless}) and also putting them and the chosen quantities $M$ and $R_\textrm{halo}$ into previous condition, one can find the relation between two parameters, which are undefined yet - particle mass $m$ and self-interaction constant $\lambda/8\pi$:
\begin{equation}
\label{eq:Rcondition}
    \frac{m}{eV}=10^{-24}\sqrt{1.27+0.23 \sqrt{\pi^3+1.98\cdot10^{100}\frac{\lambda}{8\pi} } }
\end{equation}

Next, we need to choose the rest of the parameters
We assume the particle mass to be $m = 3\cdot10^{-24}\text{eV}$, which has approximately the same order as particle masses used in different BEC DM simulations, and find from the  Eq.(\ref{eq:Rcondition})  the corresponding self-interaction constant
$\lambda/8\pi=5.6\cdot10^{-98}$. 
Further, we use the fixed mass $m$ and self-interaction constant $\lambda$; however, the total halo mass $M$ is not necessarily needed to be equal $M=3\cdot 10^{11}M_{\odot}$. We can tune this mass by changing the normalization constant $N_0$. 
Thus, the physical parameters, which are used for the dimensionless version of Eq.(\ref{GPP1}): $L_* = 6.35\cdot10^{19}\text{m}$, $\Omega_*^{-1} = 2.04\cdot10^{14}\text{s}$, $\epsilon =7.05\cdot10^{50}\text{J}$. 

In the case of the vortex solitons ($s>0$) the total energy Eq.(\ref{Etot}) depends on two variational parameters: $(R,\eta)$. Further details of the variational procedure are discussed for case $s=1,2,3$ in  Appendix \ref{app:VA}. 

The next step is to find a pair of variational parameters which minimizes the total energy. It can be done by solving the system of the equations:
\begin{equation*}
    \frac{\partial E}{\partial R}=0, 
    \frac{\partial E}{\partial \eta}=0.
\end{equation*}
This system is transcendental; therefore, it can be solved numerically. The pair of parameters $(R,\eta)$ is found for different values of normalization constant $N_0$. Inserting this into Gaussian ansatz Eq.(\ref{ansatz}), one can calculate the total energy of the system and chemical potential for different values of $N_0$. The results in dimensional quantities are shown in Fig. \ref{FIG:Energy_mu}. 
The effective spatial scales can be defined as follows:
\begin{equation}
R_{\text{eff}}^2 = \frac{1}{N_0}\int r_\perp^2 |\psi|^2 d\textbf{r}
    \label{eq:R_eff}
\end{equation}
\begin{equation}
Z_{\text{eff}}^2  
= \frac{1}{N_0}\int z^2 |\psi|^2 d\textbf{r}.
\label{eq:Z_eff}
\end{equation}
In case $s>0$ the definition of $R_{\text{eff}}$ differs from the one in case $s=0$ Eq.(\ref{eq:R0_eff}).
The variational analysis results for these quantities are shown in Fig. \ref{FIG:RZ_eff} as dashed lines. 




\subsection{Numerical modeling}\label{sec:Stationary_numerical}
We solve numerically the set of Eq.(\ref{eq:GPEP_num}) of nonlinear  equations using the
stabilized relaxation procedure similar to that employed in \cite{PRE05}. 


%
The fundamental soliton ($s=0$) corresponds to a spherically symmetric solution
$$\Psi(\mathbf{r},t) = \psi(r)e^{{-i\mu t}}.$$
In this case Eq.(\ref{eq:GPEP_num}) takes form
\begin{eqnarray}\label{eq:Phi_s0}\nonumber
  \mu\psi  = -\frac12\left( \frac{d^2\psi}{dr^2}+\frac{2}{r}\frac{d\psi}{dr}\right)+\left(\Phi+\psi
  ^2\right)\psi, \\
  \frac{d^2\Phi}{d r^2}+\frac{2}{r}\frac{d\Phi}{d r}=\psi^2.
\end{eqnarray}
Boundary conditions are
$\psi'(0)=0,$ and $\psi\to 0,$ at $r\to\infty$.
Gravitational potential $\Phi(r)$ for fixed spherically-symmetric condensate density distribution can be found analytically as follows:
\begin{equation}\label{eq:Phi_0}
    \Phi(r)=-\frac{\mathcal{M}_0(r)}{r}+\mathcal{D}_0(r)-\mathcal{D}_0(\infty),
\end{equation}
where 
\begin{equation}\label{eq:M_0}
    \mathcal{M}_0(r)=\int_0^r  \psi^2(\xi) \xi^2 d\xi,
\end{equation}
\begin{equation}\label{eq:D_0}
    \mathcal{D}_0(r)=\int_0^r \psi^2(\xi) \xi  d\xi.
\end{equation}
The boundary-value problem for $\psi$ has been solved numerically in coordinate space using stabilized relaxation method described in Ref. \cite{PRE05}.

For $s>0$ stationary state has cylindrical symmetry
$$\Psi(\mathbf{r},t) = \chi(r_\perp,z)e^{{is\theta}}e^{{-i\mu t}}.$$
Consider $\chi=\chi(r_\perp,z)$ and $ \Phi=\Phi(r_\perp,z)$ we obtain 
\begin{eqnarray}\label{eq:stationaryGPE_s}\nonumber
\mu \chi = -\frac12\left[\Delta_\perp^{(s)}+\frac{\partial^2}{\partial z^2}\right]\chi+\left(\Phi+\chi
  ^2\right)\chi \\
\left[\Delta_\perp^{(0)}+\frac{\partial^2}{\partial z^2}\right]\Phi  = \chi^2,
\end{eqnarray}
where $\Delta_\perp^{(s)}=\frac{\partial^2}{\partial r_\perp^2}+\frac{1}{r_\perp}\frac{\partial}{\partial r_\perp}-\frac{s^2}{r_\perp^2}$
the boundary conditions for the vortex soliton profile are
\begin{equation*}
\chi(0,z)=0;
\lim_{r_\perp\to\infty}\chi(r_\perp,z)=0;
\lim_{z\to\pm\infty}\chi(r_\perp,z)=0.
\end{equation*}
For the potential $\Phi$ we used the boundary condition Eq.(\ref{eq:Phi_0}) assuming that the potential is given by Coulomb potential with reasonable accuracy well apart of localized condensate cloud.

Using Fourier transform for $z$ coordinate we obtain from 
Eq.(\ref{eq:stationaryGPE_s}) the boundary value problem  for each Fourier harmonic. This radial problem has been solved by the stabilized iterative procedure similar to the spherically-symmetric problem described above.

The results obtained with both numerical and analytical methods are in good agreement with each other:  density functions for chosen total halo mass for cases $s=0,1,2,4$ are shown in Fig. \ref{FIG:Rho}; the energy and the chemical potential for cases $s=0,1,2,3,4$ are shown in Fig. \ref{FIG:Energy_mu}; effective radius and height are shown in Fig. \ref{FIG:RZ_eff}.
%




\section{\label{sec:dynamics}Dynamics}
The long-lived CDM structures which survive at cosmological time scales can play a crucial role in the formation and evolution of the galaxies. Thus, it is very important to verify
whether obtained steady-state solutions are stable. 
We have studied the stability of the vortex structures by direct
simulations of the propagation dynamics of perturbed vortex
solitons by applying the split-step Fourier method to solve
Eqs.(\ref{eq:GPEP_num}) numerically. The details of the numerical procedure are discussed in Appendix \ref{app:numerical}.

The dynamical simulations of $s$-charged DM structures were initiated with the perturbed steady-state wave function $\Psi_s$ of the form  $\Psi|_{t=0}=\Psi_s \cdot \left[1+\varepsilon \cdot \cos(L\theta)\right]$, where $\varepsilon$ is the perturbation amplitude and integer $L$ corresponds to the azimuthal symmetry of perturbation.
\begin{figure}[htpb]
	\includegraphics[width = 3.4 in]{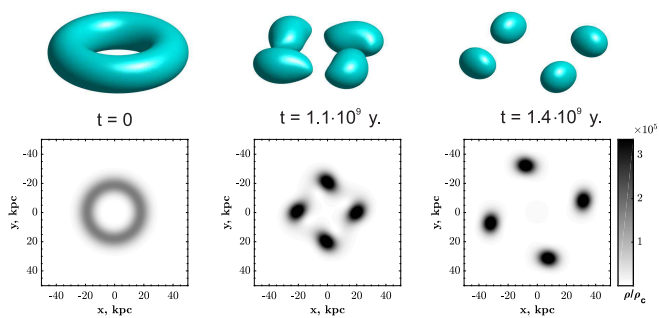}
	\caption{The snapshots of the 3D isosurface of the condensate density (upper row) and the normalized condensate density in $(x,y)$ plane (lower row) for $s=4$. 
	The snapshots are given for three indicated moments of time. Note that $s=4$ vortex disintegrates into four flying away fragments keeping the kinetic energy of the vortex flow.}
	\label{fig:dynamics_m=4}
\end{figure}

\begin{figure}[htpb]
	\includegraphics[width = 3.4in]{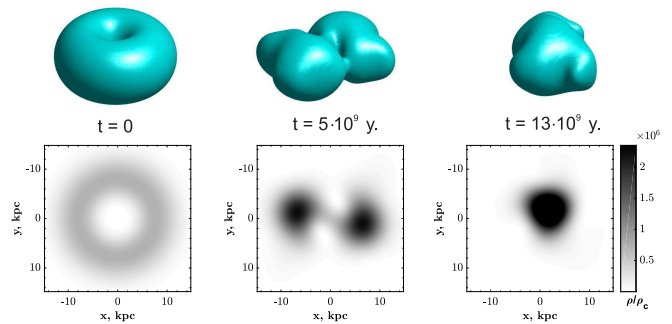}
	\caption{The same as in Fig. \ref{fig:dynamics_m=4} for $s=2$. Note that a doughnut-shaped vortex transforms into a single-connected blob with complex condensate flow at the periphery of the galactic halo. }
	\label{fig:dynamics_m=2}
\end{figure}

\begin{figure}[htpb]
    	\includegraphics[width = 3.4in]{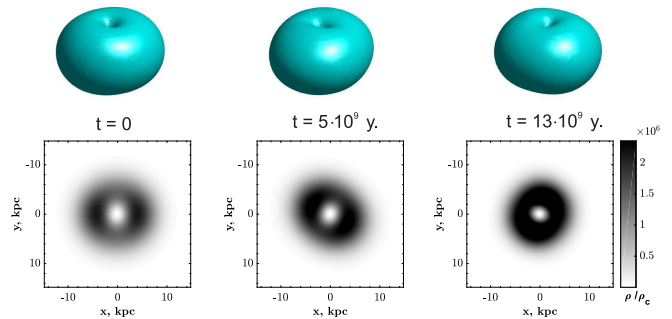}
	\caption{The same as in Fig. \ref{fig:dynamics_m=4} for stable single-charged ($s=1$) DM vortex soliton. It is remarkable that even being strongly perturbed ($\varepsilon=0.1$, $L=2$) the vortex survives during the Universe lifetime.}
	\label{fig:dynamics_m=1}
\end{figure}

Evolution of the condensate density for different topological charges is illustrated in Figs. \ref{fig:dynamics_m=4}, \ref{fig:dynamics_m=2}, and \ref{fig:dynamics_m=1}. 
We have found that DM halo with embedded multi charged $s>1$ vortex is unstable due to azimuthal symmetry-breaking instability. It is remarkable that vortex structures with $s\ge 4$ disintegrate into the filaments taking away the kinetic energy of the condensate vortex superflow (see Fig. \ref{fig:dynamics_m=4}). Using simple estimates based on the conservation total energy it is straightforward to find that for the galactic halo of mass $M=3\cdot 10^{11} M_{\odot}$  kinetic energy of the vortex flow dominates the gravitational binding energy for $s\ge 3$. Thus even $s=3$ vortex in principle can disintegrate into flying away filaments according to this estimate. However, we never observed the disintegration of the vortex states with $s<4$ in our numerical simulations. The vortex solitons with $s=2,3$ are also unstable, but the initial doughnut-shaped DM halo transforms into a single-connected blob with vortex flow located mostly at the periphery of the halo (see an example of such evolution for $s=2$ in Fig. \ref{fig:dynamics_m=2}).



With no surprise, we observed stable evolution of the fundamental soliton ($s=0$), which exhibit periodic oscillation of the width and amplitude caused by initial perturbation. It is much more remarkable that the single-charged $s=1$ vortex soliton appears to be stable even being strongly perturbed. Stable evolution of the DM vortex  is illustrated in  Fig. \ref{fig:dynamics_m=1} for $L=2$ azimuthal perturbation having the amplitude $\varepsilon=0.1$. 



\section{Conclusions}\label{sec:conclusions}
We have studied superfluid self-gravitating BEC with nonzero angular momentum. We have analyzed stationary three-dimensional vortex soliton states with different topological charges. By means of analytical variational analysis, we predict the main features of steady vortex soliton solutions, which are in good agreement with our numerical results. 

Using direct numerical simulations of the (3+1)D Gross-Pitaevski-Poison system we studied the evolution of spinning DM cloud with typical galactic halo mass and radius. We have found that while all multi-charged vortex states ($s\ge2$) are unstable, a single-charged vortex soliton ($s=1$) and fundamental soliton ($s=0$) are extremely robust and survive during the lifetime of the Universe.
In the strict sense, even quite robust dynamics for a huge time does not prove rigorously stability of the DM structure. In the present work, we restrict stability analysis fixing the DM halo mass by a typical value. This raises the question of whether azimuthal instability is suppressed for $s=1$ DM vortices with an arbitrary mass, or there is a stability threshold for spinning galactic halo formed by superfluid BEC.   Further investigations including linear stability analysis are needed for a severe test of stability. 

 A comprehensive analysis of the interactions between spinning superfluid DM and luminous matter is beyond the scope of the present work. This problem merits
a separate study, that is now in progress, and the results will be published elsewhere. Nevertheless, some tentative general conclusions from our theoretical results can be made. 
Both outcomes with stable vortex solution $(s=1)$ and vortex decay $(s>1)$ provide interesting results that might have a connection to galaxy structures. One shows that for unstable $s=2$ and $s=3$  CDM structures vortices transfer from the centre to the periphery of the halo, which might be related to the galaxy rotation curve problem. We have found that vortices with $s\ge 4$ are unstable to decay into fragments, which constrain from above the angular momentum of the considered CDM structures.  The other, for stable $s=1$ vortex CDM structures, one can assume that the baryonic matter can gather in the central region of the galaxies, following the analogy to atomic BEC and thermal atoms filling vortex threads.  We hope that research on this topic can shed a light on the problem of the formation of a supermassive black hole, which is seen at the centre of almost every large galaxy. 

 Novel, increasingly accurate observational evidence combined with essential progress in theoretical and computational methods are promising in terms of confirming, constraining or discarding the superfluid model of CDM in the nearest future. 
 We believe that the results, described in the present work,  will help to elucidate important properties of dark matter, which is a problem of fundamental interest.

\section{Acknowledgments}
This work was supported by National Research Foundation of Ukraine through grant No. 2020.02/0032. The authors are grateful to Artem Oliinyk for discussions and comments about this paper.

\appendix


\section{\label{app:VA}Variational analysis, energy for $s \geq 1$}
Here we present the details of the variational results for vortex solitons.
It appears, that the result of integration Eq.(\ref{eq:dimless_energy}) when $\eta>1$ differs from the one when $0<\eta<1$.  
For $s=1$ dimensionless total energy in each case is as follows:
\begin{widetext}
\begin{equation*}
   E^{s=1}/\epsilon = \frac{N_0(1+4\eta^2) }{4R^2\eta^2}+\frac{N_0^2}{8\sqrt{2}\pi^{3/2}R^3\eta} +W^{s=1}/\epsilon,
\end{equation*}
where the last term is gravitational energy of interaction and it is  sensitive to the value of $\eta$:
\begin{equation*}
 W^{s=1}_{\eta>1}/\epsilon =  -\frac{N_0^2}{\sqrt{2}\pi^{3/2}R}\frac{3\eta(1-2\eta^2)\sqrt{\eta^2-1}+(11-24\eta^2+16\eta^4)\text{arccosh}(\eta)}{64(\eta^2 - 1)^{5/2}}
 \end{equation*} 
\begin{equation*}
W^{s=1}_{\eta<1}/\epsilon =  -\frac{N_0^2}{\sqrt{2}\pi^{3/2}R}\frac{3\eta(1-2\eta^2)\sqrt{1-\eta^2}+(11-24\eta^2+16\eta^4)\text{arctan}(\sqrt{1/\eta^2-1})}{64(1-\eta^2)^{5/2}}
  \end{equation*} 

For $s=2$ dimensionless total energy in each case is as follows:
\begin{equation*}
    E^{s=2}/\epsilon = \frac{N_0(1+6\eta^2) }{4R^2\eta^2}+\frac{3N_0^2}{32\sqrt{2}\pi^{3/2}R^3\eta}+W^{s=2}/\epsilon,
\end{equation*}
    where the last term is gravitational energy of interaction and it is  sensitive to the value of $\eta$:
    \begin{equation*}
    \begin{split}
     W^{s=2}_{\eta>1}/\epsilon=&-\frac{N_0^2}{\sqrt{2}\pi^{3/2}R}\frac{\eta\sqrt{\eta^2-1}(201-794\eta^2+1080\eta^4-592\eta^6)}{4096(\eta^2-1)^{9/2}}-\\
     &-\frac{N_0^2}{\sqrt{2}\pi^{3/2}R}\frac{ \big[ 585+32\eta^2(-77+126\eta^2-96\eta^4+32\eta^6) \big] \text{arccosh}(\eta)}{4096(\eta^2-1)^{9/2}}
    \end{split}
 \end{equation*}

\begin{equation*}
    \begin{split}
    W^{s=2}_{\eta<1}/\epsilon=&-\frac{N_0^2}{\sqrt{2}\pi^{3/2}R}\frac{\eta\sqrt{1-\eta^2}(201-794\eta^2+1080\eta^4-592\eta^6)}{4096(1-\eta^2)^{9/2}}-\\
    &-\frac{N_0^2}{\sqrt{2}\pi^{3/2}R}\frac{\big[585+32\eta^2(-77+126\eta^2-96\eta^4+32\eta^6)\big]\text{arctan}(\sqrt{1/\eta^2-1})}{4096(1-\eta^2)^{9/2}}
    \end{split}
\end{equation*}

For $s=3$ dimensionless total energy in each case is as follows:
\begin{equation*}
    E^{s=3}/\epsilon = \frac{N_0(1+8\eta^2) }{4R^2\eta^2}+\frac{5N_0^2}{64\sqrt{2}\pi^{3/2}R^3\eta}+W^{s=3}/\epsilon,
\end{equation*}
    where the last term is gravitational energy of interaction and it is  sensitive to the value of $\eta$:
    \begin{equation*}
    \begin{split}
     W^{s=3}_{\eta>1}/\epsilon=&\frac{N_0^2}{\sqrt{2}\pi^{3/2}R}\frac{\eta\sqrt{\eta^2-1}(2\eta^2-1)\big[3147+8\eta^2(-1546+2521\eta^2-1960\eta^4+736\eta^6)\big]}{65536(\eta^2-1)^{13/2}}+\\
     &+\frac{N_0^2}{\sqrt{2}\pi^{3/2}R}\frac{\Big[-8267+51480\eta^2-16\eta^4\big\{8469+64\eta^2(-189+156\eta^2-72\eta^4+16\eta^6)\big\}\Big]\text{arccosh}(\eta)}{65536(\eta^2-1)^{13/2}}
    \end{split}
 \end{equation*}

\begin{equation*}
    \begin{split}
    W^{s=3}_{\eta<1}/\epsilon=&\frac{N_0^2}{\sqrt{2}\pi^{3/2}R}\frac{\eta\sqrt{1-\eta^2}(2\eta^2-1)\big[3147+8\eta^2(-1546+2521\eta^2-1960\eta^4+736\eta^6)\big]}{65536(1-\eta^2)^{13/2}}+\\
     &+\frac{N_0^2}{\sqrt{2}\pi^{3/2}R}\frac{\Big[-8267+51480\eta^2-16\eta^4\big\{8469+64\eta^2(-189+156\eta^2-72\eta^4+16\eta^6)\big\}\Big]\text{arctan}(\sqrt{1/\eta^2-1})}{65536(1-\eta^2)^{13/2}}
    \end{split}
\end{equation*}
\end{widetext}

It is noteworthy that the results for different $\eta$ can be achieved using analytic continuation of the function $\text{arctan}(\sqrt{1/\eta^2-1})$ in region, where $\eta>1$ and vice versa. We have calculated analytically total energy for the case $s=4$ as well, but the results are too cumbersome to be presented here.



\section{\label{app:numerical}Numerical method for dynamical simulations}


Here we present details of the numerical methods used for dynamical simulations in our work. For a recent review of the numerical methods used for modelling self-gravitating BECs see \cite{2020PhyD..40332301P}. 
There are two different types of numerical methods to deal with the partial differential equations with the Laplacian term. One is to use a finite difference scheme, determining the value of the Laplacian at each point of the grid. Another possibility is to compute the Laplacian in Fourier space, while the other terms in coordinate space. This is accomplished by implementing the split-step Fourier method (SSFM), which profits from the efficiency of the fast Fourier transform (FFT) algorithm. 
In this case, zero boundary conditions for $\Phi$ can be convenient in preventing the influence of its periodical structure. 
Using FFT at each time step we have solved the Helmholtz equation:
\begin{equation}\label{eq:Helholtz}
    \Delta\Phi(\mathbf{r},t)=|\Psi(\mathbf{r},t)|^2+\alpha^2\Phi,
\end{equation}
which is free of the mathematical singularity of the Coulomb potential in Fourier space.
The parameter $\alpha$, responsible for screening, has been chosen so that the potential $\Phi(\textrm{r},t)$ of the dynamical GPP fits the solution of the Poisson equation $\Phi(\textrm{r})$ for the stationary state in a region of the high density. Note that the potential of the stationary GPP has been solved numerically with no screening, as described in Sec. \ref{sec:Stationary_numerical}.
To fit the amplitude of the potential even better we used additional normalizing parameter $\beta$ as follows: 
${\Phi_0}=\beta {\Phi}$, where $\Phi$ is the solution of Eq.(\ref{eq:Phi_s0}) for $s=0$ and Eq.(\ref{eq:stationaryGPE_s}) for $s>0$.
In our simulations we choose the parameters $\alpha$ and $\beta$ for each initial condition to obtain an appropriate correspondence between exact  potential and approximate screened potential in the region where condensate density $|\psi|^2$ has significant support.

\bibliography{references}

\end{document}